**Title** Optically interrogated liquid crystal-based, charge-mode accelerometer telemetry

**Zourab Brodzeli[1,2*], Can Nerse[1], Benjamin Halkon[1], John Canning[2,3,4], Sebastian Oberst[1*]**

[1]Centre for Audio, Acoustics and Vibration, School of Mechanical and Mechatronic Engineering, Faculty of Engineering and Information Technology, University of Technology Sydney, 123 Broadway, Ultimo, NSW 2007, Australia
[2]PhoenixZ Pty Ltd, Suite 305, 30 Kingsway, Cronulla, NSW 2230, Australia
[3]Photonics & Optical Commun. School of Electrical Engineering & Telecommunications, University of New South Wales, Sydney, NSW 2052, Australia
[4]School of Mathematics and Physics, University of South China, 28 Changsheng West Road, Hengyang City, Hunan Province, China

Address correspondence to: Zourab Brodzeli; zourab.brodzeli@phoenix-z.com and Sebastian Oberst sebastian.oberst@uts.edu.au

## Abstract

A hybrid optoelectronic vibration sensing system is presented, combining a conventional charge-mode piezoelectric accelerometer with an optical fiber interface via a liquid crystal-based electrical-to-optical transducer. This configuration allows the accelerometer's electrical output to be converted into an optical signal, enabling remote signal transmission over optical fiber. The same fiber could also be used to supply power to the system, offering a compact and interference-resistant solution suitable for challenging environments such as those with high voltages or explosive risks. Experimental results demonstrate a displacement resolution ranging from $\Delta z = 0.015$ nm at 0.160 nm displacement, up to $\Delta z = 0.720$ nm at 351 nm displacement, or equivalently, from $\alpha = 0.0005\ \mathrm{m{\cdot}s^{-2}}$ to $0.057\ \mathrm{m{\cdot}s^{-2}}$ in acceleration. Despite undergoing electrical-to-optical and optical-to-electrical signal conversion, the system maintains high sensitivity, low noise levels, and signal integrity. This proof-of-concept optically interrogated accelerometer highlights a path toward reducing sensor infrastructure requirements by eliminating the need for conventional power supplies and electromagnetic shielding, which are typically required in fully electrical accelerometer systems.

## INTRODUCTION

An accelerometer senses acceleration and can be used to measure vibrations of a structure [1,2]. Accelerometers are widely used in both industrial and scientific applications, such as navigation systems [3,4], touchdown and acceleration monitoring for aircraft and missiles [5,6], vibration-based fault detection in rotating machinery [7,8], and image stabilization in cameras and display systems [9,10]. In many of these scenarios, especially those involving large or complex mechanical systems, vibration or displacement must be measured at multiple locations. This requirement has led to the development of distributed point-sensing networks using accelerometers [11,12], creating a network of devices. The physical connectivity of this network becomes an important element of the overall system device.

Traditional accelerometer networks rely on copper wiring to connect each sensor to a multi-channel data acquisition system or interrogator. However, copper wires introduce several limitations: high electrical resistance, limited cable length, and significant vulnerability to electromagnetic interference (EMI). These challenges can degrade signal quality, limit network coverage, and increase system complexity. Furthermore, copper wiring poses safety risks due to electrical propagation, making it unsuitable for use in explosive or flammable environments such as coal mines, oil refineries, gas pipelines, wood-processing facilities, or grain mills [13,14]. These limitations create serious obstacles when deploying sensor equipment, like terrestrial mining drills, in such hazardous settings. Consequently, there is a growing demand for alternative accelerometer systems that can operate safely and effectively in harsh environments, without incurring the high infrastructure costs associated with specialized safety-certified designs.

In this work, a hybrid sensor approach that addresses both, network and telemetry constraints by integrating a conventional charge-mode accelerometer with an optically interrogated, liquid crystal-based optical fiber transducer is presented [15,16]. Unlike copper-based systems, optical telemetry offers a range of advantages: immunity to EMI, intrinsic safety, support for large-area coverage (tens of square kilometers), high bandwidth, and reliable data transmission. Moreover, the hybrid optically interrogated accelerometer (OIA) system offers a simpler and potentially more cost-effective alternative to fully optical fiber-based vibration sensors, and conventional accelerometers can be used for high-temperature, high-frequency, or high-shock environments where micro-electromechanical systems (MEMS) devices would fail.

Since the 1980s, several generations of optical fiber vibration sensors have emerged, including designs that incorporate MEMS based on Fabry–Perot interferometers (FPIs) [17-20] and fiber Bragg gratings (FBGs) [21,22]. These high-sensitivity systems rely on optical resonant cavities: FPIs utilize a Fabry–Perot micro-cavity in which changes in cavity length cause shifts in the frequency of the reflected interrogation signal, while FBGs employ distributed, laser-inscribed weak but coherent etalons that cumulatively reflect a narrow wavelength band. In FBG-based FPIs, the reflected wavelength varies in response to both mechanical length and potentially out-of-plane changes and refractive index variations caused by vibration. While these sensors offer high sensitivity[1], their performance is often limited by the complexity and stability requirements of the packaging and optical measurement system [23,24]. Sensors based on fiber Bragg gratings (FBGs) rely on detecting shifts in their resonant (Bragg) wavelength in response to external stimuli such as strain, temperature, or vibration. To accurately measure these shifts, high-resolution optical interrogation systems are required, typically optical spectrum analyzers (OSAs) or tunable laser-based demodulation units. These instruments must resolve small changes in wavelength, often on the order of picometers, which demands exceptional spectral resolution and stability. High-resolution OSAs capable of this precision are expensive, bulky, and may be impractical for large-scale or field-deployable sensing systems. These costs and complexity limit the widespread adoption of FBG-based sensors in applications where simpler or more rugged systems are preferred.

In contrast, the hybrid approach avoids reliance on spectral and thus phase-sensitive mechanisms that are costly and challenging to measure with high resolution and stability. Unlike traditional FBG-based systems that require expensive, high-resolution OSAs to detect spectral shifts, the hybrid optical approach relies on direct intensity modulation. Specifically, the amplitude of light reflected from the transducer varies linearly with the mechanical displacement of the sensing element, effectively converting acceleration into an optical intensity signal. This signal is then detected using simple, low-cost photodetectors instead of OSAs. Because the proprietary transducers exhibit excellent linearity, having a total harmonic distortion of -35 dB, the output signal accurately reflects the physical input with minimal distortion. This design reduces system complexity and cost while maintaining high fidelity, making it suitable for scalable and field-deployable sensing applications.

[1] The sensitivity of the Sinocera accelerometer connected with the in-line amplifier is around 1,700 mV/g. (measuring range 600g pk).

Instead, the sensor's electrical output is translated directly into optical intensity variations, which are simpler and more robust to detect. This eliminates the need for high-precision phase tracking, improving reliability and reducing system cost compared to traditional FBG or FPI-based sensors. Unlike these traditional fiber sensors, which require complex and expensive equipment to track tiny changes in light wavelength or phase, our hybrid sensor system works in a much simpler way. Instead of measuring how the color (wavelength) of the light changes, which requires high-resolution, very stable, and costly optical instruments, we convert the sensor's signal into changes in the brightness (intensity) of light. The amplitude can be easily measured using low-cost photodetectors. Avoiding the need for precise wavelength or optical phase shifts makes the sensing technology more reliable in harsh environments and simplifies system construction.

For network deployment, wavelength division multiplexing (WDM) FBGs can still be incorporated into the telemetry system for wavelength-division multiplexing (WDM), enabling multiple sensors to be identified and interrogated over a single fiber. By leveraging off-the-shelf piezoelectric accelerometers and integrating them into a flexible optical network, the method extends and enhances the capabilities of current established sensing technologies, supporting both remote operation and scalable, distributed deployment in challenging environments [27].

## METHODS

**Experimental methodology.** The optically interrogated accelerometer (OIA) shown in Fig. 1 is a self-contained system developed from a commercial piezoelectric accelerometer (CA-YD-127, Sinocera Piezotronics). Its output is routed through an audio amplifier (BOB09816, SparkFun Electronics), which drives an optical fiber–pigtailed liquid crystal transducer [15]. In this setup, the electrical signal from the accelerometer is first amplified, then converted into an optical signal by the transducer. This conversion is accomplished using a liquid crystal variable reflectance mirror, whose reflectivity changes in response to the applied voltage. Reflectance is controlled by the orientation of liquid crystal molecules, which is sensitive to changes in both voltage and polarization. As a result, the reflected optical signal is polarized along the liquid crystal axis to reduce insertion loss. Remote optical interrogation is performed using a polarized optical source and photodetector, with transmission and reception occurring via standard single-mode telecommunications optical fiber (SMF28) [24].

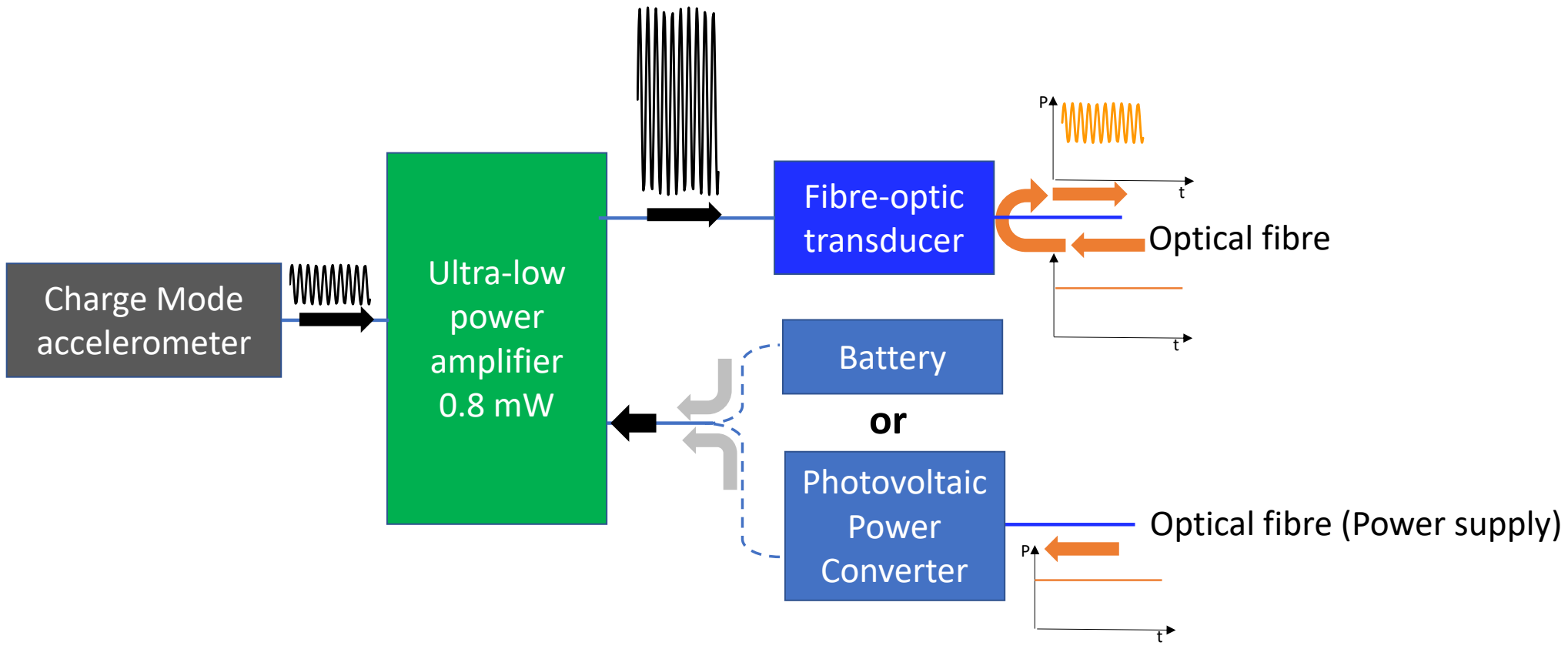

**Fig. 1.** Schematic of our optically interrogated accelerometer (OIA). The charge mode accelerometer generates an electrical output proportional to the applied acceleration. This output is amplified by an ultra-low-power-consuming amplifier. The electrical output of the amplifier is transformed into optical by the optical fiber transducer. The ultra-low power-consuming amplifier can be powered either by a battery or by power-over-fiber.

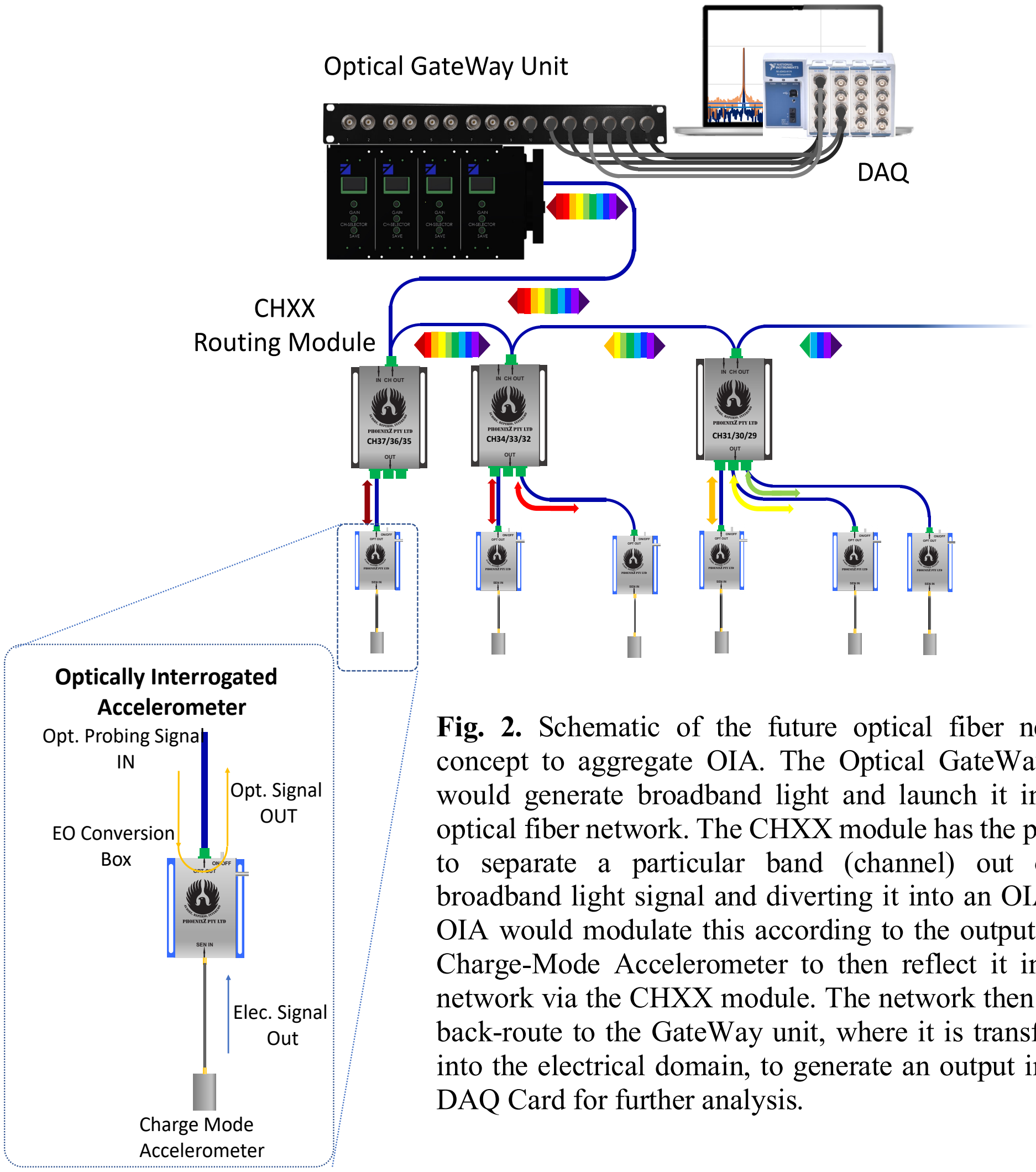

**Fig. 2.** Schematic of the future optical fiber network concept to aggregate OIA. The Optical GateWay Unit would generate broadband light and launch it into the optical fiber network. The CHXX module has the purpose to separate a particular band (channel) out of the broadband light signal and diverting it into an OIA. The OIA would modulate this according to the output of the Charge-Mode Accelerometer to then reflect it into the network via the CHXX module. The network then would back-route to the GateWay unit, where it is transformed into the electrical domain, to generate an output into the DAQ Card for further analysis.

The operating power required for the accelerometer and amplifier stages is $P$ = 0.8 mW, which is low power compared to a conventional in-line charge accelerometer. This can be supplied locally by a rechargeable battery or photovoltaic cell, or by light delivered via optical fiber and converted into electrical energy using power over fiber [17,19,25] methods, including a photovoltaic power converter, rendering the device optical.

The charge-mode accelerometer generates a voltage proportional to the applied force. The electrical power of the transducer is low, ranging from $P$ = 0.05 to 400 nW. As the liquid crystal rotates under varying applied voltage, its reflectivity (level of distortion THD = -35 dB) changes linearly with voltage [15]. The device shown in Fig. 1 can operate at ultralow powers with minimum switching at $P \sim 0.27$ nW, which is optimized to be in the range required for switching the optical fiber transducers. The power consumption of a typical ultra-low power-consuming audio amplifier is in the range of a few mW (for model BOB09816 $\Delta P$ = 0.8 mW). These low operating powers enable the possibility of powering the device by light, delivered to the component by optical fiber and converted into electrical energy by photovoltaic cells.

This OIA requires a specific optical fiber network as demonstrated in Figure 2. A gateway unit is a device that acts as a bridge between sensors and the rest of the system. It collects data, converts signals if necessary, and transmits the information to locations where it can be stored, displayed, or acted upon. In this context, the gateway unit consists of an optical interrogation light source and a photodetector. Routing Modules (CHXX) that control how data and signals are directed through the system, a data acquisition (DAQ) card, and the OIA modules. This network utilizes wavelength division multiplexing (WDM) to transmit multiple channels through a single optical fiber, allowing multiple sensors to be connected along the same optical fiber and operate simultaneously. The Gateway unit generates broadband light and launches it into the network. The CHXX routing module separates a specific wavelength, or WDM channel, from the broadband signal and directs it to the OIA, while the remaining wavelength channels continue through the network to other modules and sensors. The OIA modulates the incident light in response to acceleration and reflects the encoded signal to the Gateway unit via the optical network. The Gateway unit separates the sensor wavelengths, directing each to its respective photodetector, where the optical signals are converted back into electrical signals. These electrical signals are then read by the DAQ card for further processing. In the optical domain, transmission distances can span several kilometers and can be extended even further with optical fiber amplification, demonstrating the advantage of this hybrid approach.

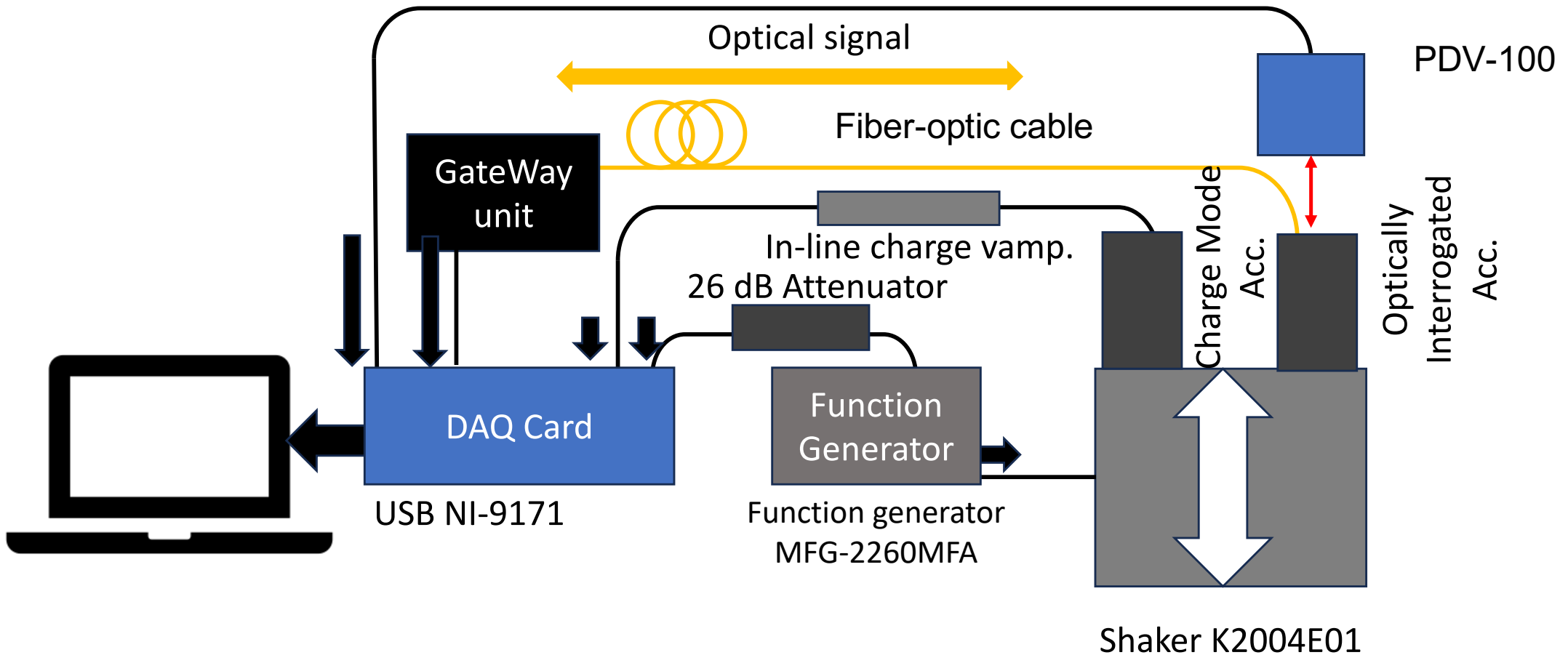

**Fig. 3.** Experimental setup for optically driven excitation and vibration signal acquisition. A function generator drives a light source (yellow path), which emits modulated optical signals through a fiber to the excitation point (red dot) on the vibrating structure. The response is measured via a photodetector and converted into an electrical signal by the GateWay unit, which then feeds into a DAQ card for data acquisition and processing. Bi-directional black arrows indicate signal and control communication; yellow arrows represent optical pathways; grey boxes represent physical components such as amplifiers and sample mounts. The large white arrow denotes the direction of vibrational excitation.

### Commercial benchmark and measurement systems used

As shown in Figure 3, a single-input/single-output vibration test protocol has been established to measure the response of the OIA. The converted output was compared against two independent vibration monitoring sensing methods. The first was a direct electrical measurement using a commercial charge-mode accelerometer (CA-YD-127, Sinocera Piezotronics), also with an in-line charge amplifier (4705M4, Dytran Instruments). The second was a free-space optical measurement using a laser Doppler vibrometer (PDV-100, Polytec), which interrogated the optical fiber-coupled charge accelerometer. Both independent methods were used to measure the displacement of the shaker, on which the accelerometers were mounted, when a voltage signal was applied, and their results were compared to the measurements obtained from the OIA. Further, we compared the performance of the OIA with that of the original CA-YD-127 accelerometer. A photograph of the experimental setup is shown in Figure 4.

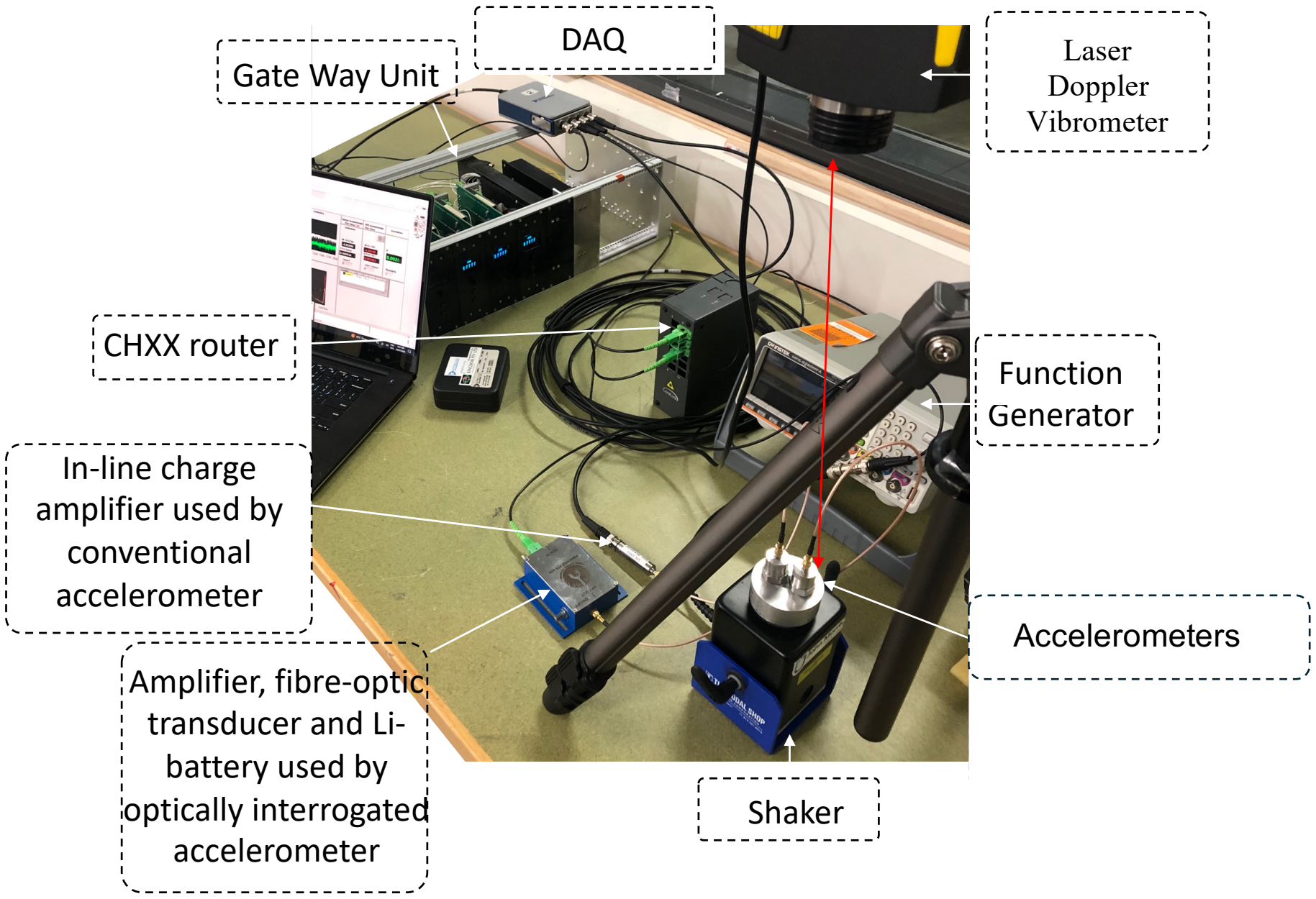

**Fig. 4.** Photograph of the setup to characterize the optically interrogated. The system includes a trip-mounted PDV focused on the aluminum holder, which is optically excited using a fiber-coupled source. The excitation is modulated via a function generator and controlled through the GateWay unit and DAQ system connected to a laptop for data acquisition.

## RESULTS

Figure 5(A,B) shows a fast Fourier transform of the electrical output from the OIA mounted on the shaker, in response to a 1 kHz alternating current signal with an amplitude of 0.5 mV. The physical displacement of the shaker driven by the applied voltage corresponds to $\Delta z$ = 0.7 X $10^{-3}$ µm. Figure 5 (C,D) shows the output of the accelerometer as well as the amplified charge mode accelerometer and laser Doppler vibrometer, for various shaker displacements at a constant frequency of 1 kHz.

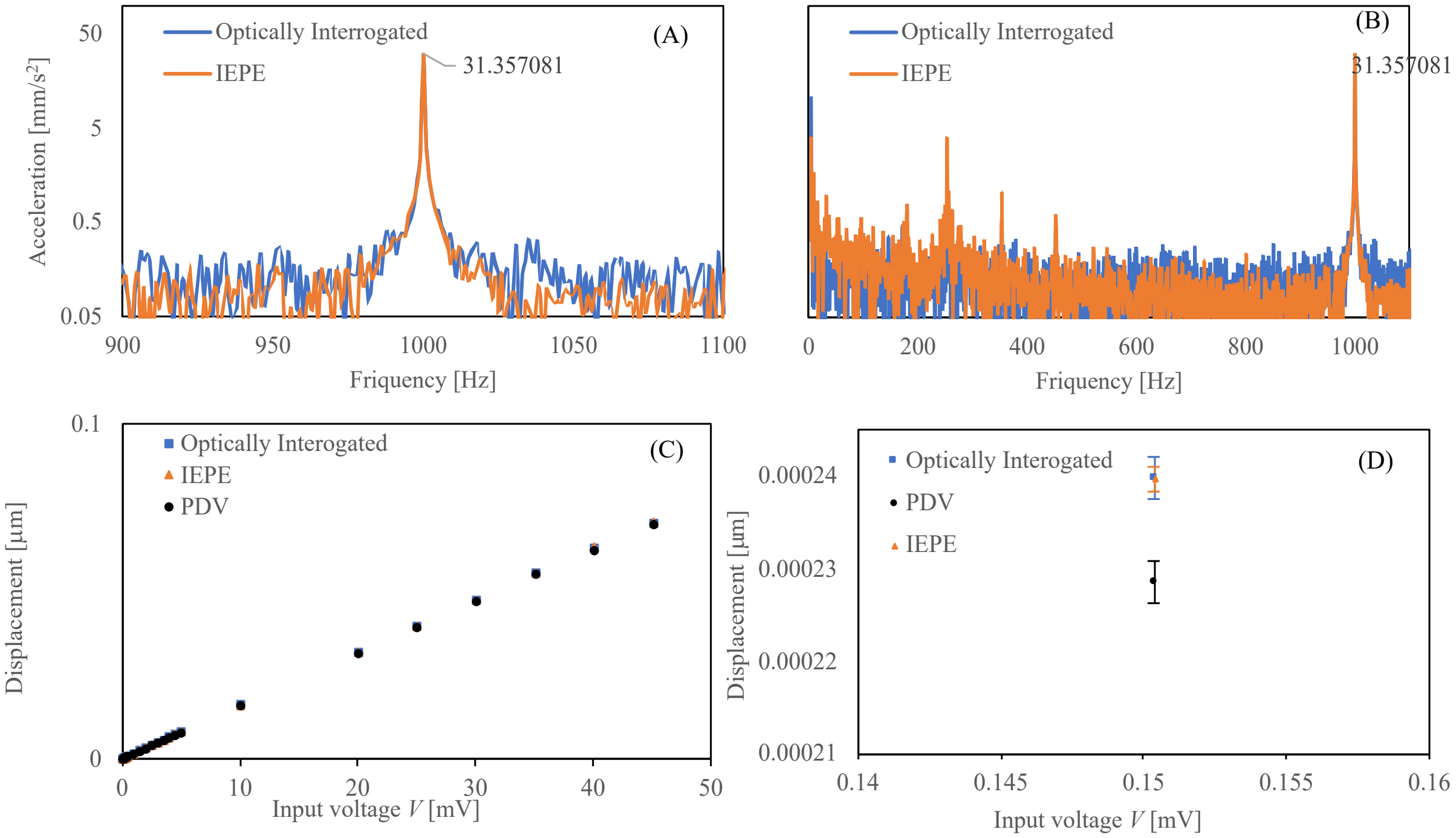

**Fig. 5.** (a) FFT of the OIA output at $v$ = 1 kHz, acceleration $a$ = 31.4 mm/s$^2$ and displacement of the shaker of $\Delta z$ = 0.7 nm, with applied shaker $V$ = 0.5 mV The output of three accelerometers (the OIA, the charge mode accelerometer with in-line charge amplifier, and the laser Doppler vibrometer), recalculated as displacement for different applied voltage to the shaker at constant $v$ = 1 kHz; (b) The same figure zoomed in to highlight the errors of the measurement for each sensor. This error corresponds to the standard deviation of each sensor's measurements during two minutes of sampling at 1 Hz.

The resolution of the measurements from the OIA, the charge-mode accelerometer, and the PDV were evaluated. For each voltage level shown in Figure 5(C), 120 data points were collected while maintaining a constant amplitude and frequency of the voltage applied to the shaker. The resulting standard deviation ($\sigma$) of these measurements is shown in Figure 6. From this data, the resolution of each sensor for different displacements was determined, then summarized and compared (Table 1).

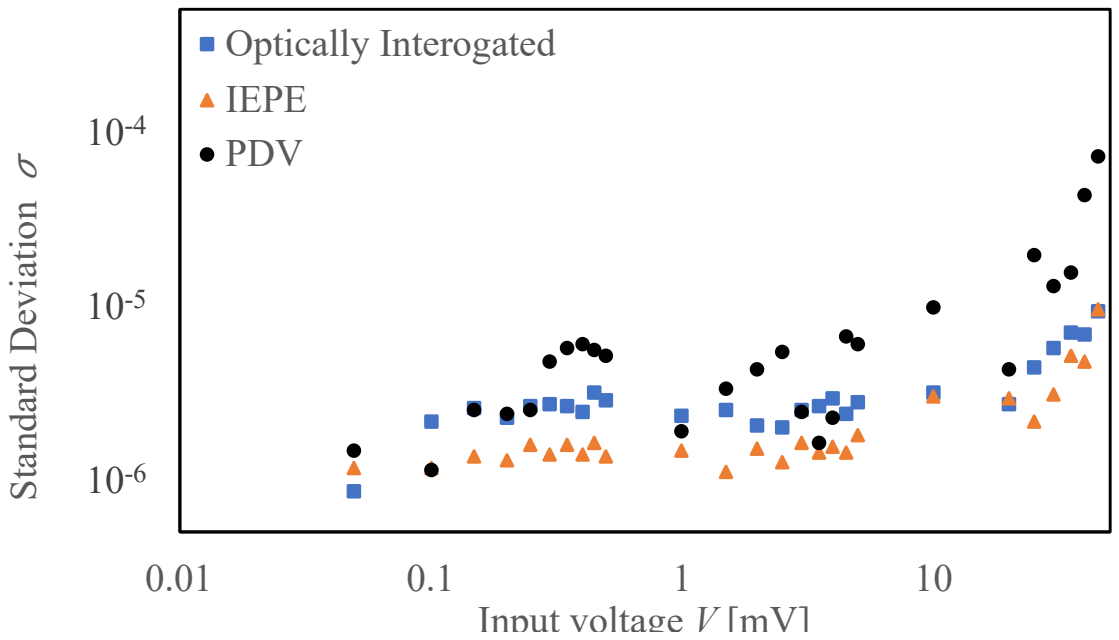

**Fig. 6.** Analysis of the resolution of the measurement of the OIA, charge mode accelerometer and laser Doppler vibrometer: Each data point represents the standard deviation of 120 measurements of the same displacement at frequency of the shaker of 1 kHz. The displacement was varying with voltage applied to the shaker.

**Table 1.** This table represents a numerical description of Fig. 6 of the output of the OIA, the IEPE accelerometer and the PDV at nine different displacements, *z*, at shaker *ν* = 1 kHz. Those measurements were used to determine the vibration resolution of the OIA, which varies with the displacement of the shaker.

| OIA | | | IEPE | | | PDV | | |
|---|---|---|---|---|---|---|---|---|
| *z* [µm] | *σ* | % | *z* [µm] | *σ* | % | *z* [µm] | *σ* | % |
| 0.00016 | 2.5E-06 | 1.57 | 0.00016 | 1.1E-06 | 0.70 | 0.00015 | 2.5E-06 | 1.61 |
| 0.0016 | 2.5E-06 | 0.15 | 0.0016 | 1.5E-06 | 0.09 | 0.0015 | 3.3E-06 | 0.21 |
| 0.016 | 2.7E-06 | 0.01 | 0.016 | 2.9E-06 | 0.01 | 0.015 | 4.3E-06 | 0.02 |

The frequency response of the charge-mode accelerometer was compared to that of the OIA using a linear sine sweep from 10 Hz to 2 kHz over 60 s. The results reveal a noticeable discrepancy between the two sensor outputs in the low-frequency range (50–450 Hz). This deviation is attributed to differences in signal conditioning: the conventional charge-mode accelerometer uses an in-line charge amplifier with a tailored gain profile to compensate for reduced sensitivity at low frequencies, while the optically interrogated system relies on a standard audio amplifier without such compensation.

## DISCUSSION

This work presents a hybrid sensing approach that bridges traditional charge-mode accelerometers with fiber-optic telemetry by employing a liquid crystal-based electrical-to-optical conversion mechanism. In doing so, we successfully demonstrate that a conventional piezoelectric accelerometer, when coupled to a liquid crystal-based transducer, can transmit good-quality acceleration data optically with minimal power consumption and even without requiring electromagnetic shielding or local electrical infrastructure. The system achieves a resolution as fine as 0.015 nm at a displacement of 0.16 nm and demonstrates consistent measurement accuracy across a broad displacement range, up to 351 nm. The linearity and robustness of the OIA signal indicate that this method maintains high sensitivity and reliability while avoiding common limitations of copper-based wiring, particularly in environments prone to electrical interference or explosion hazards.

Compared to previous efforts to develop all-optical or interferometric vibration sensing technologies, such as FBG or FPI systems [17–22], this approach offers the advantage of a simpler optical readout based on intensity modulation, eliminating the need for phase-tracking. This reduces cost, eases implementation, and improves mechanical stability. Although these optical systems often achieve higher theoretical resolution, they require complex optical alignment and spectral demodulation equipment. In contrast, the hybrid system offers a practical compromise, retaining the advantages of remote optical telemetry while utilizing reliable, readily available piezoelectric sensing elements.

Nevertheless, certain limitations merit discussion. The observed discrepancy in the frequency response between the OIA and the conventional charge-mode accelerometer, particularly in the 50 Hz to 450 Hz range, underscores the importance of proper signal conditioning. The found deviation is attributable to the difference in amplification strategies: a conventional accelerometer uses a charge amplifier with a frequency-dependent gain profile, whereas the OIA currently relies on a generic audio amplifier. While this can be addressed with a custom-designed amplifier tailored to the accelerometer's characteristics, such refinements were beyond the scope of this proof-of-concept demonstration.

The ability of this hybrid system to function with power-over-fiber (PoF) infrastructure, potentially eliminating the need for local batteries or power supplies, opens important applications in distributed structural health monitoring, mining, aerospace, and other safety-critical domains. This also aligns with recent developments in optically powered sensor platforms [17,19,26], which underscore the feasibility of powering entire sensor networks using light transmitted over kilometers of optical fiber [27-29].

This study demonstrates the feasibility of integrating conventional electromechanical sensors into modern optical telemetry frameworks, creating a pathway for safer, simpler, and more scalable sensing systems. By combining legacy sensor hardware with liquid crystal-based optical interfacing, this hybrid approach offers a compelling solution for vibration sensing in future distributed instrumentation networks in harsh environments. For very low power solutions, it would be interesting to extend the system with MEMS accelerometers for power levels which can be in the order of μW.

To improve the overall performance of the system, further engineering optimization of the amplifier, transducer, and fiber interface is essential to increase sensitivity and broaden the operational bandwidth. A comparative study against MEMS-based optical accelerometers could also yield valuable benchmarking data and inform future design improvements.

Although wavelength-division multiplexing (WDM) was included in the conceptual design (see Fig. 2), it has not yet been implemented in the current prototype. Future developments will need to integrate WDM to enable multiplexed sensing and optical power delivery through shared fiber pathways, an essential step towards demonstrating the system's scalability.

Finally, to validate the system's real-world applicability, large-scale environmental testing is planned, particularly in high-electromagnetic interference (EMI) or hazardous zones. Such trials are critical to establish proof-of-concept and confirm robust field deployment capabilities, including for applications in large-scale infrastructure monitoring [28,29].

## ACKNOWLEDGMENTS

**Funding:** The research conducted was supported by the Australian Research Council under the Linkage Project funding scheme (project No. LP200301196) and the Discovery Project funding scheme (project No. DP200100358).

**Author contributions:** Z.B., C.N., B.J., and S.O. conceptualized the experiments. Z.B., C.N. and B.J. collected the data. Z.B. and J.C. developed the methodology. Z.B. and S.O wrote the initial manuscript version. S.O. acquired the funding, managed, and supervised the project. All authors contributed to the data interpretation, discussion of the results, and the manuscript revision. Correspondence and requests for materials should be addressed to Z.B. and S.O.

**Competing interests:** The first author (ZB) is the CEO and PZ which has submitted a provisional patent application 2022902934 with CN and SO as co-inventors. Otherwise, the co-authors declare that they have no known competing interests or personal relationships that could have appeared to influence the work reported in this paper.

**General:** The first author acknowledges assistance and discussions with Drs Shahrokh Sepehrirahanama and Ha Pham.

## DATA AVAILABILITY

The authors declare that the primary data supporting the findings of this study are available within the paper. Additional data are available from the corresponding authors upon request.